\documentclass[aps,prl,twocolumn,superscriptaddress,amsmath,amssymb]{revtex4}
\usepackage{float}
\usepackage{makeidx}
\usepackage{graphicx,amsmath}
\usepackage{CJK}
\usepackage{colordvi}
\usepackage{color}
\usepackage{amssymb}
\usepackage{epstopdf}
\usepackage{bm}
\usepackage{graphicx,epsfig}
\usepackage{mathrsfs}
\usepackage{dcolumn}
\usepackage{natbib}
\usepackage{miniplot}
\usepackage{subfigure}
\begin{document}
\begin{CJK*}{GB}{gbsn}

\title{Experimental observation of one-dimensional superradiance lattices in ultracold atoms }

\author{Liangchao Chen}

\affiliation{State Key Laboratory of Quantum Optics and Quantum
Optics Devices, Institute of Opto-Electronics, Shanxi University,
Taiyuan 030006, P.R.China }

\affiliation{Collaborative Innovation Center of Extreme Optics,
Shanxi University, Taiyuan 030006, P.R.China}

\author{Pengjun Wang$^*$}

\affiliation{State Key Laboratory of Quantum Optics and Quantum
Optics Devices, Institute of Opto-Electronics, Shanxi University,
Taiyuan 030006, P.R.China }

\affiliation{Collaborative Innovation Center of Extreme Optics,
Shanxi University, Taiyuan 030006, P.R.China}

\author{Zengming Meng}

\affiliation{State Key Laboratory of Quantum Optics and Quantum
Optics Devices, Institute of Opto-Electronics, Shanxi University,
Taiyuan 030006, P.R.China }

\affiliation{Collaborative Innovation Center of Extreme Optics,
Shanxi University, Taiyuan 030006, P.R.China}

\author{Lianghui Huang}

\affiliation{State Key Laboratory of Quantum Optics and Quantum
Optics Devices, Institute of Opto-Electronics, Shanxi University,
Taiyuan 030006, P.R.China }

\affiliation{Collaborative Innovation Center of Extreme Optics,
Shanxi University, Taiyuan 030006, P.R.China}

\author{Han Cai$^\dagger$}

\affiliation{Interdisciplinary Center of Quantum Information and Department of Physics, Zhejiang University, Hangzhou 310027, P.R.China}
\affiliation{Institute of Quantum Science and Engineering, Texas A$\&$M University, College Station, TX 77843, US}

\author{Da-Wei Wang}

\affiliation{Interdisciplinary Center of Quantum Information and Department of Physics, Zhejiang University, Hangzhou 310027, P.R.China}

\author{Shi-Yao Zhu}

\affiliation{Interdisciplinary Center of Quantum Information and Department of Physics, Zhejiang University, Hangzhou 310027, P.R.China}
\affiliation{Synergetic Innovation Center of Quantum Information and Quantum Physics,
University of Science and Technology of China, Hefei, Anhui 230026,
P. R. China}

\author{Jing Zhang$^{\ddagger}$}

\affiliation{State Key Laboratory of Quantum Optics and Quantum Optics Devices,
Institute of Opto-Electronics, Shanxi University, Taiyuan 030006,
P.R.China }

\affiliation{Synergetic Innovation Center of Quantum Information and Quantum Physics,
University of Science and Technology of China, Hefei, Anhui 230026,
P. R. China}

\begin{abstract}
We measure the superradiant emission in a one-dimensional (1D)
superradiance lattice (SL) in ultracold atoms. Resonantly excited
to a superradiant state, the atoms are further coupled to other
collectively excited states, which form a 1D SL. The directional
emission of one of the superradiant excited states in the 1D SL is
measured. The emission spectra depend on the band structure, which
can be controlled by the frequency and intensity of the coupling
laser fields. This work provides a platform for investigating the
collective Lamb shift of resonantly excited superradiant states in
Bose-Einstein condensates and paves the way for realizing higher
dimensional superradiance lattices.

\end{abstract}
\maketitle
\end{CJK*}

Ultracold atoms have been a highly
controllable system for investigating condensed matter physics \cite{Greiner2002} and
providing versatile applications in quantum optics and quantum information
processing \cite{Bloch2008}. In addition to absorption imaging \cite{Davis1995},
light scattering is a common method for observing and analyzing
physics in cold atoms, such as the optical Bragg scattering
of cold atomic gases
\cite{Birkl1995, Weidemuller1995, Slama2005, Schilke2011} and the antiferromagnetic correlations
in a Fermi-Hubbard model \cite{Hart2015}.
Different from the optical Bragg scattering with hot atoms
\cite{Bajcsy2003,Bae2008,Zhang2011,Wang2013}, highly coherent atomic matter
waves are usually involved in cold atoms \cite{Meystre}. Superradiance \cite{Dicke1954} is also important
in studying collective effect in atoms \cite{Feld1973,DeVoe1995,Kimble2015}. In Bose-Einstein condensates (BEC)
off-resonant pumping fields are usually used, such as the BEC superradiance in a free space \cite{Inouye1999} and
the superradiance phase transition in an optical cavity \cite{Baumann2010}. Dipolar interactions induced by off-resonant light in BEC can result in rotons \cite{Dell2003} and Cooper pairs \cite{Zhang1993}, which have a substantial effect on the excitation spectra of the BEC \cite{Steinhauer2002}. Off-resonant fields are also used to investigate the collective effect in the superradiance of noncondensed cold atoms \cite{Araujo2016,Roof2016}.

BEC superradiance with on-resonant pumping is more difficult to observe
since the pumping field and the superradiant radiation are in the
same direction. However, many interesting phenomena in
superradiance and subradiance of the phase-correlated timed Dicke
states \cite{Scully2006, Bienaime2012, Scully2015, Bromley2016,
Goban2015, Guerin2016} and the collective Lamb shift
\cite{Scully2009, Rohlsberger2010, Keaveney2012, Meir2014}
only happen when the superradiant states are near or on-resonantly
excited. Recently, the superradiance lattice composed by timed Dicke
states in momentum space was proposed in noncondensed atoms
\cite{Wang2015}. Since the direction of the emission of the superradiant states
in the superradiance lattice can be different from that of the probe field, we
are able to investigate the on-resonantly excited superradiant states in a BEC.
In this Letter, we report the experimental realization of the
superradiance lattice in a BEC of $^{87}$Rb ultracold atoms. The light
emitted by one of the superradiant states in the superradiance
lattice is measured. Different from the
well-known Bragg scattering of cold atoms with periodically
modulated density \cite{Schilke2011}, the spectra of the superadiant
emission demonstrate the band structure of a one dimensional tight-binding lattice.

To highlight the underlying physics, we consider a three-level $\Lambda$-type
system with one excited state $|e\rangle$ and two ground states
$|g\rangle$ and $|m\rangle$.
The atoms are initially prepared in a BEC of the ground state $|g\rangle$.
We approximate this state as $|G\rangle\equiv|N,\mathbf{q}=0\rangle_g$ which means
$N$ atoms in the state $|g\rangle$ with zero momentum $\mathbf{q}$. This is a good
approximation for the current investigation although the atoms are
in a trap and the ground state contains other momentum components.
A single photon resonantly absorbed by the BEC results in the state
$|1,\mathbf{k}_p\rangle_e|N-1,0\rangle_g$ where $\mathbf{k}_p$ is the
wavevector of the probe photon, i.e., one atom is excited from $|g\rangle$
to $|e\rangle$ and acquires a recoil momentum $\mathbf{k}_p$. We first analyze
the spontaneous emission of this excited state. The interaction Hamiltonian between
the atoms and the vacuum modes is ($\hbar=1$)
\begin{equation}
H_1=\sum\limits_{\mathbf{k},\mathbf{q}}g_\mathbf{k}a^\dagger_\mathbf{k}
b^\dagger_{g}(\mathbf{q-k})b_{e}(\mathbf{q})+H.c.,
\end{equation}
where $a_\mathbf{k}$ is the annihilation operator of the photon with
wavevector $\mathbf{k}$, $g_\mathbf{k}$ is the coupling constant
between the photon and atoms, $b_{i}(\mathbf{q})$ is the annihilation operator
of the atoms in the internal state $|i\rangle$ with momentum $\mathbf{q}$.

One particular vacuum mode with wave vector $\mathbf{k}_p$ dominates the spontaneous emission in the excited BEC.
It is easy to find that the coupling between the single photon excited state $|1,\mathbf{k}_p\rangle_e|N-1,0\rangle_g$
and the ground state is
\begin{equation}
\langle 0, 0|_e \langle N,0|_gH_1|1,\mathbf{k}_p\rangle_e|N-1,0\rangle_g=\sqrt{N}g_{\mathbf{k}_p} a^\dagger_{\mathbf{k}_p},
\end{equation}
accompanied by the radiation of a photon with wave vector $\mathbf{k}_p$. The interaction is enhanced by
$\sqrt{N}$ times, which is a signature of the superradiance. Instead, if the
atoms radiate a photon with momentum $\mathbf{k}\neq \mathbf{k}_p$,
the coupling strength is
\begin{equation}
\langle 0, 0|_e \langle 1, \mathbf{k}_p-\mathbf{k}; N-1,0|_g H_1|1,\mathbf{k}_p\rangle_e|N-1,0\rangle_g=g_\mathbf{k}a^\dagger_{\mathbf{k}}.
\end{equation}
Here $\left\langle 1, \mathbf{k}_p-\mathbf{k}; N-1,0\right|_g$ denotes the final state bra with one atom with momentum $\mathbf{k}_p-\mathbf{k}$ and $N-1$ atoms with zero momentum in the state $|g\rangle$. Therefore, the spontaneous decay of the state $|1,\mathbf{k}_p\rangle_e|N-1,0\rangle_g$
is dominated by the superradiance emission in the mode $\mathbf{k}_p$. The state $|1,\mathbf{k}_p\rangle_e|N-1,0\rangle_g$ is
a BEC version of the timed Dicke state
$|e_{\mathbf{k}_p}\rangle=\frac{1}{\sqrt{N}}\sum_j e^{i\mathbf{k}_p\cdot\mathbf{r}_j}|g_1, g_2, ..., e_j, ..., g_N\rangle$ \cite{Scully2006}, where
$|i_j\rangle$ ($i=e,g$) and $\mathbf{r}_j$ are the internal states and position of the $j$th atom.

\begin{figure}
\centerline{
\includegraphics[width=8cm]{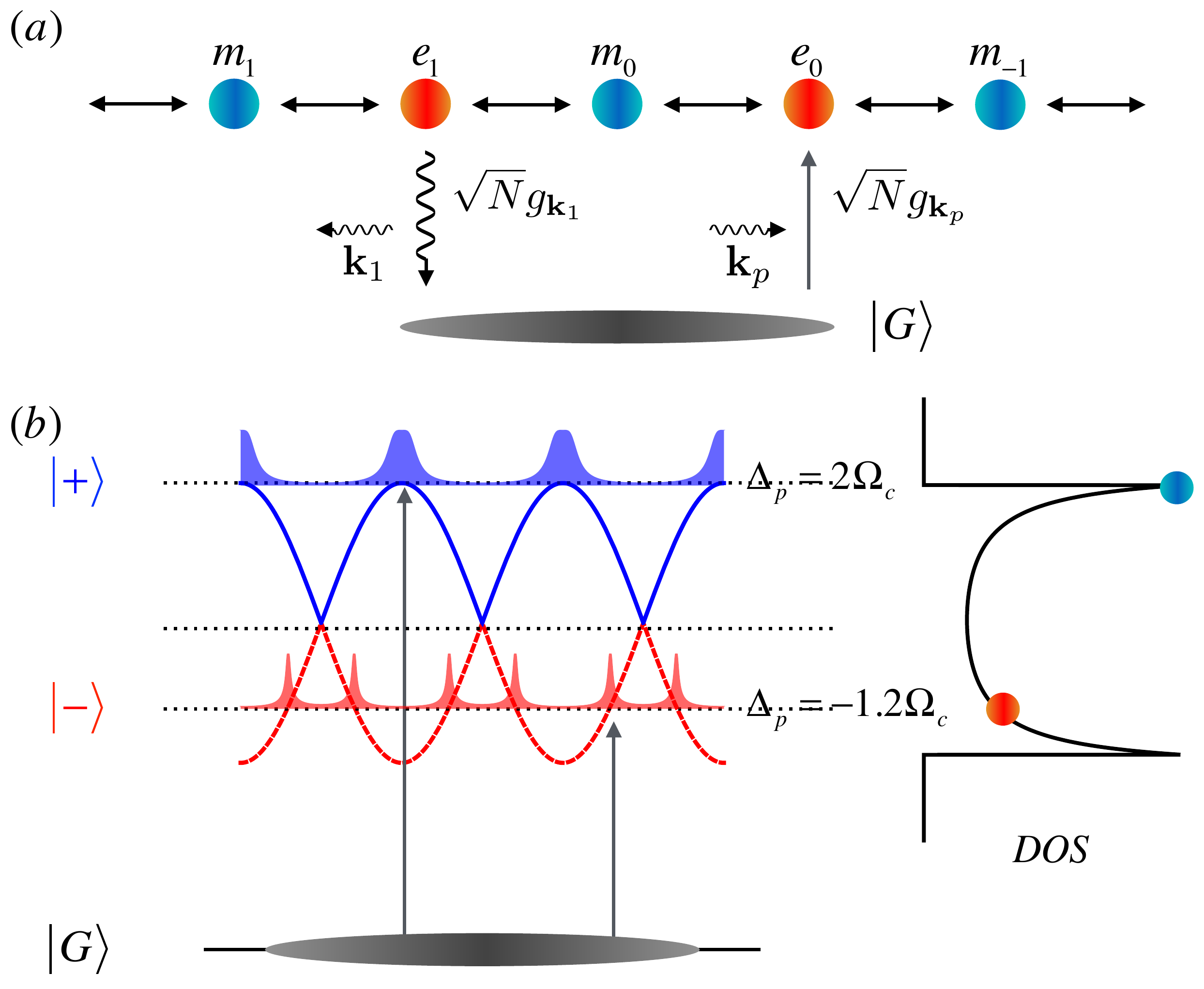}} \vspace{0.1in}
\caption{\textbf{Band structure of the superradiance lattice in BEC.}(a) In the superradiance lattice,
the pump laser populates $|e_0\rangle$, which is a superradiant state and coupled to other collective states. Among these states,
only $|e_1\rangle$ is superradiantly coupled to the ground states and its directional emission
is measured. (b) The atoms dressed by the standing wave
have different energies of their eigenstates $|\pm\rangle=(|e\rangle\pm|m\rangle)/\sqrt{2}$ at
different positions. The excitation probabilities of the BEC atoms at different positions
is plotted for two different pump frequencies. The total excitation probabilities are proportional
to the density of states of a tight-binding model of collectively excited states.
\label{scheme} }
\end{figure}

\begin{figure}
\centerline{
\includegraphics[width=6cm]{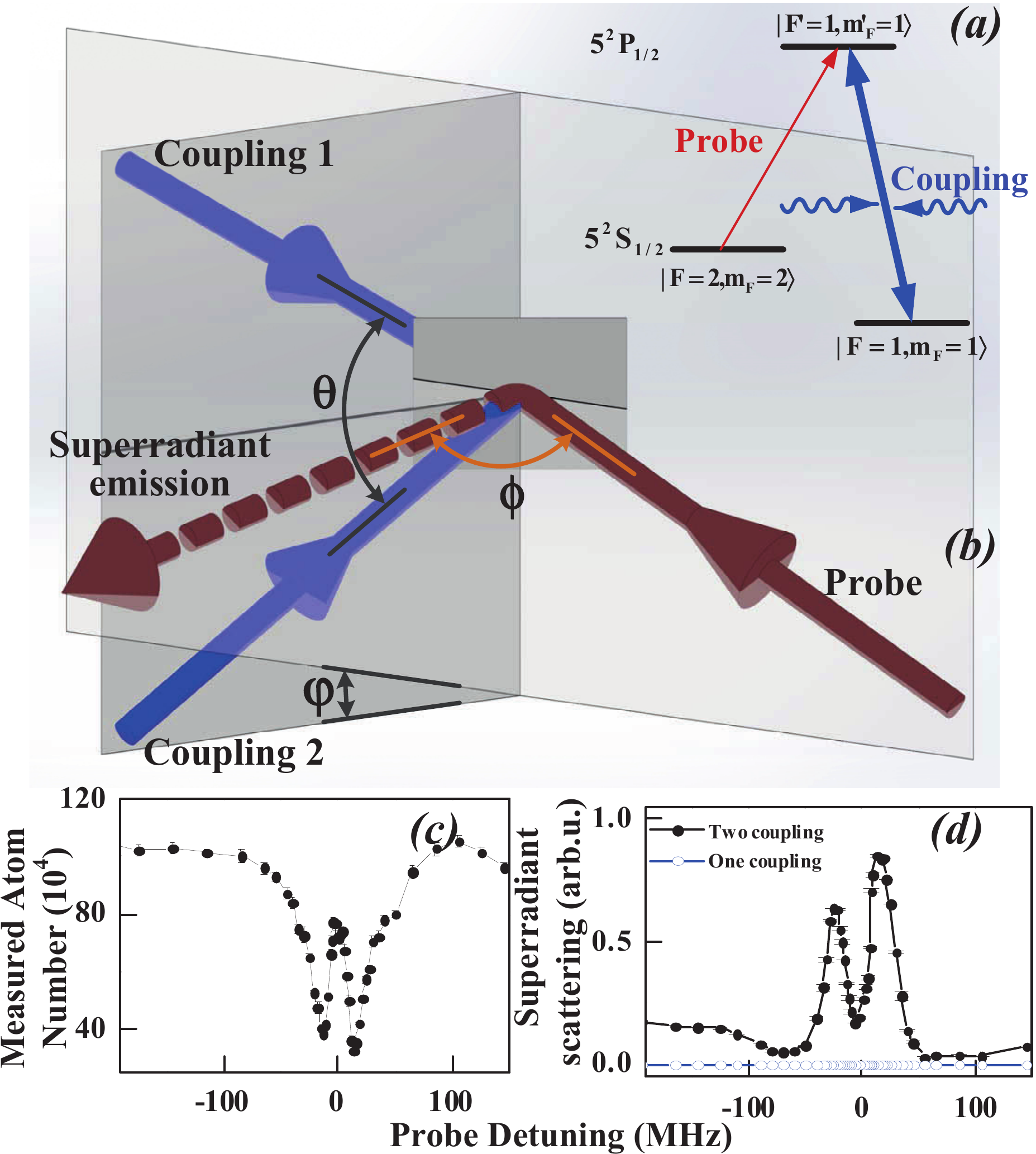}} \vspace{0.1in}
\caption{\textbf{Experimental configuration of the 1D superradiance lattice in BEC.} (a) Energy diagram of the
$5^{2}S_{1/2}-5^{2}P_{1/2}$ transition of $^{87}$Rb. (b) The experimental
geometry and the laser configuration. The three planes in the figure are the plane
of the two coupling beams, the plane of the probe-superradiant beams, and the equal intensity
plane of the coupling beams. (c) EIT spectrum by measuring the remaining
atoms with only one coupling laser. The power of
each coupling laser is 200 $\mu W$ and on resonance ($\Delta_{c}=0$). The power of the
probe laser is 25 $\mu W$. (d) Intensity of the superradiant scattering with
a pair of coupling lasers. Each of the coupling laser is 200 $\mu W$
and the other parameters are the same as in (c). The data points are simply connected. \label{experiment} }
\end{figure}

However, such a radiation is difficult to observe in experiments since the excitation and radiation signals
are in the same direction. In order to observe the directional radiation, we need to change the wavevector
of the excited state. This can be done by introducing two coherent fields that
drive the transition between the excited state $|e\rangle$ and another ground state $|m\rangle$.
The interaction Hamiltonian is
\begin{equation}
H_c=\sum\limits_{\mathbf{q},j}\Omega_j b^\dagger_{m}(\mathbf{q-k}_{cj})b_{e}(\mathbf{q})+H.c.,
\label{Hc}
\end{equation}
where $\Omega_j$ are the Rabi frequencies of the coupling fields with the wavevectors $\mathbf{k}_{cj}$ ($j=1,2$).
We introduce a short-hand notation
$|e_l\rangle\equiv|1,\mathbf{k}_l\rangle_e|N-1,0\rangle_g$ and $|m_l\rangle\equiv|1,\mathbf{k}_l-\mathbf{k}_{c1}\rangle_m|N-1,0\rangle_g$ with $\mathbf{k}_l=\mathbf{k}_p-l(\mathbf{k}_{c1}-\mathbf{k}_{c2})$ and $l$ an integer. Through the Hamiltonian in Eq. (\ref{Hc}), $|e_l\rangle$ and
$|m_l\rangle$ form a 1D tight-binding chain, as shown in Fig.~\ref{scheme} (a). Considering the on-site potential induced by the detuning of the coupling field $\Delta_{c}\equiv\nu_c-\omega_{em}$ with $\nu_c$ the coupling field
frequency and $\omega_{em}$ the transition frequency between $|e\rangle$ and $|m\rangle$, we can write down
the interaction Hamiltonian in the subspace expanded by $|e_l\rangle$ and $|m_l\rangle$ in a tight-binding form,
\begin{equation}
\begin{aligned}
H_I=&\sum\limits_{l}\frac{\Delta_c}{2}(|m_l\rangle\langle m_l|-|e_l\rangle\langle e_l|)\\
&+(\Omega_1|m_l\rangle\langle e_l|+\Omega_2 |e_{l-1}\rangle\langle m_l|+H.c.).
\end{aligned}
\label{HI}
\end{equation}
In this lattice, if $|\mathbf{k}_f|=|\mathbf{k}_p|$ for a certain $|e_f\rangle$,
a vacuum mode with wavevector $\mathbf{k}_f$ can couple the excited state
$|e_f\rangle$ back to the ground state via directional superradiant emission in $\mathbf{k}_f$. Other states
with $|\mathbf{k}_l|\neq |\mathbf{k}_p|$ cannot find a vacuum mode to achieve a superradiant enhancement in their spontaneous emission.
The kinetic energy due to the recoil can be neglected (see Supplementary Material \cite{supp}). In our experiment, $f=1$, as shown in Fig. \ref{scheme} (a) and the radiation of $|e_{1}\rangle$ is detected.

A pure BEC with typically $5\times10^5$ $^{87}$Rb atoms is prepared
in the $|g\rangle\equiv|F=2,m_{F}=2\rangle$ hyperfine ground state
sublevel confined in a cross-beam dipole trap at a wavelength near
1064 nm. The geometric mean of the trapping frequencies is
$\overline{\omega}\simeq2\pi\times80$ Hz. The atomic size is
estimated in the Thomas-Fermi regime to be 20 $\mu m$, when the
scattering length for $|g\rangle$ state at zero magnetic field is
about 100$a_{0}$. The $D_{1}$ line (around 795 nm) of $^{87}$Rb atom
is considered as a simple three-level $\Lambda$-type model as shown
in Fig. \ref{experiment} (a) due to the large hyperfine splitting of
816.8 MHz between the two excited states. We choose the other two
relevant hyperfine levels $|e\rangle\equiv|F'=1,m'_{F}=1\rangle$,
$|m\rangle\equiv|F=1,m_{F}=1\rangle$. A homogeneous bias magnetic
field along the $z$ axis (gravity direction) is provided with
$B_{0}=0.6$ G by a pair of coils operating in the Helmholtz
configuration. A pair of strong coupling laser beams with the
intersecting angle $\theta=48^{o}$ drive the transition between
$|e\rangle$ and $|m\rangle$, as shown in Fig. \ref{experiment} (b).
The coupling laser beams have the waist ($1/e^{2}$ radius) about 280
$\mu m$ at the position of the BEC. The standing wave pattern formed by
the coupling fields have about 20 periods in the atomic gases. A
weak probe laser used to pump the atoms from $|g\rangle$ to $|e\rangle$
has a waist about 600 $\mu m$. The frequency locking and detuning of
the coupling and probe lasers are described in the Supplementary Material
\cite{supp}. The coupling and probe lasers illuminate atoms
simultaneously with 20 $\mu s$. The intersecting angle between the
superradiant emission and the probe light is about
$\phi\sim180^{o}-\theta=132^{o}$. In order to obtain the dark
background and high signal-noise ratio for detecting the
superradiant emission, the intersecting angle between the plane of
the two coupling beams and the plane of the probe-superradiant beams
is $\varphi=11^{o}$, as shown in Fig. \ref{experiment} (b). The
resulting superradiant emission is measured with an EMCCD.

If we only have one coupling laser, the two states $|e_0\rangle$ and $|m_0\rangle$ and the ground state
form an EIT configuration. Since the probe (pump) beam has a waist much larger
than the coupling beams and the size of the BEC,
we measure the remaining atoms by the time of flight
absorption image after turning on the coupling and probe pulse
for 20 $\mu s$, rather than measure the transmitted probe beam. Fig. \ref{experiment} (c) shows a transparency window at the centre.
As a comparison to the latter experiment with two coupling beams, the superradiant emission in
this case is measured to be zero, as shown in Fig. \ref{experiment} (d) (empty circles). In contrast, the solid dots in Fig. \ref{experiment} (d) show a typical superradiant emission when we have two coupling lasers. Two peaks due to the density of states (DOS) of the 1D
tight-binding lattice (see Fig. \ref{scheme} (a)) were observed.

\begin{figure}
\centerline{\includegraphics[width=6cm]{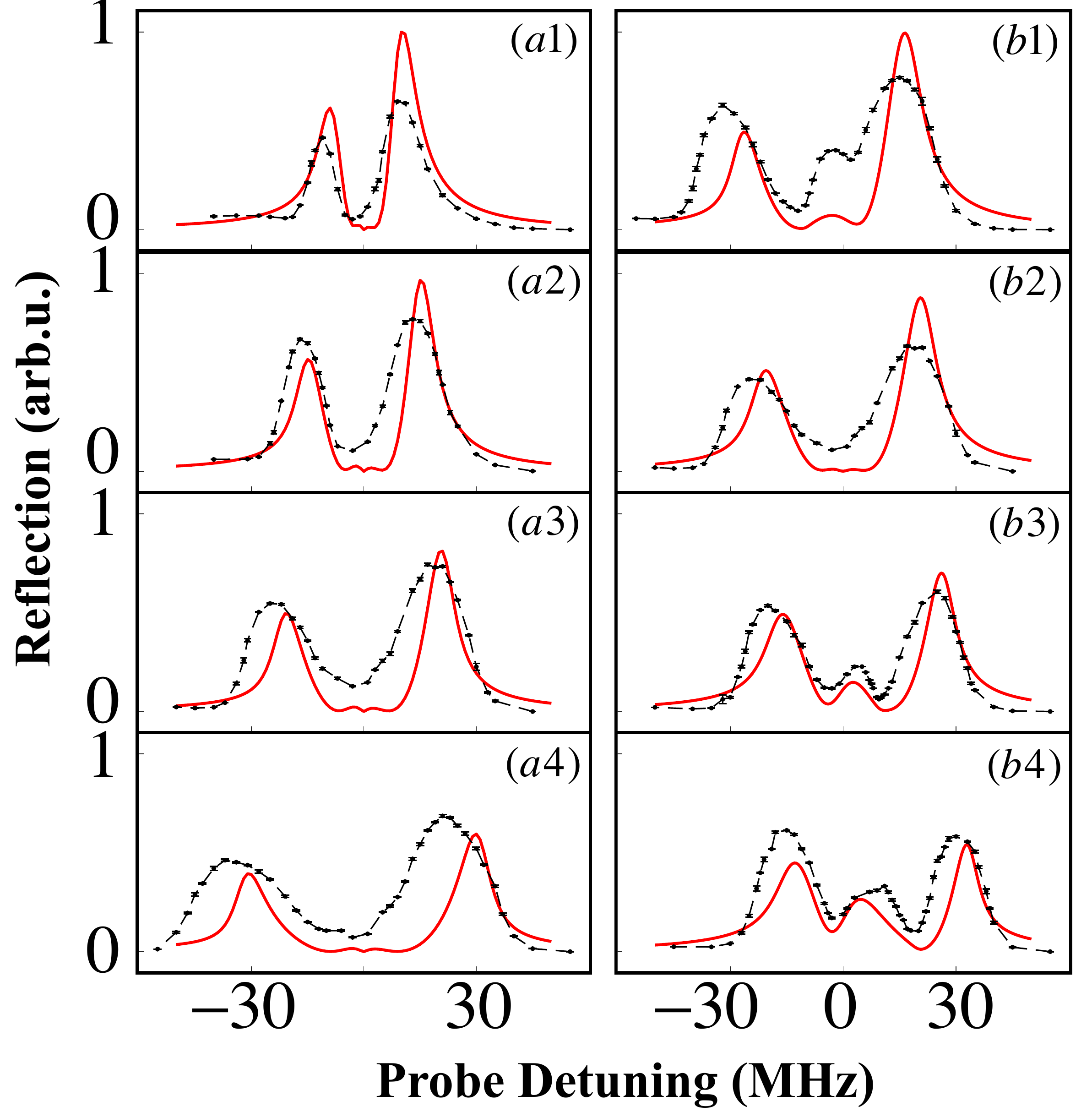}} \vspace{0.06in}
\caption{\textbf{Superradiant emission spectra for different intensities and detunings of the coupling laser.}
(a1)-(a4). The
intensity of the coupling laser are (a1) 50$\mu W$,
(a2) 100$\mu W$, (a3) 200$\mu W$, and (a4) 400$\mu W$. The
coupling laser is on resonance and the power of the probe light is 25
$\mu W$. (b1)-(b4). The detunings of the
coupling laser are (b1) $\Delta_{c}=-10$ MHz, (b2) $0$ MHz, (b3) $10$ MHz, and (b4) $20$ MHz. The powers of each coupling laser and the probe laser are 200 $\mu W$
and 25 $\mu W$. The data points are simply connected with black lines and the red curves are the theoretical fitting. \label{Fig3} }
\end{figure}

The superradiance lattice band structure can be easily tuned by the
laser parameters. For instance, the band gap can be opened by tuning
the detunings of the coupling fields from the atomic transition frequency between
$|e\rangle$ and $|m\rangle$. The bandwidth can be tuned by the laser
intensities of the coupling laser field. The superradiant emission
from the state $|e_1\rangle$ depends on this band structure. The
probability that the superradiance lattice gets excited is roughly
proportional to the DOS at the probe field frequency [as shown in Fig.
{\ref{scheme}} (a)], characterized by the two peaks at the end of
the spectra.
The excitation has a periodic structure in position
space, which means that it contains discrete momentum components.
One of these components is the superradiant excited state
$|e_1\rangle$, which inherits the two-peak feature of the DOS, as
shown for different coupling field strengths in Fig. \ref{Fig3} (a).  Since the band width
is $4\Omega_c$ where $\Omega_c=\Omega_1=\Omega_2$, the separation between the two peaks gets larger when we
increase the coupling laser power.
In Fig. \ref{Fig3} (b) we plot the emission spectra for different detunings of the coupling fields. The coupling field detuning introduces an on-site potential in the superradiance
lattice; i.e., the $|e_l\rangle$ and $|m_l\rangle$ states have an energy offset $\Delta_c$, as shown in Eq. (\ref{HI}). This opens a bandgap
and introduces asymmetry of the $|e\rangle$ components in the two bands. When $\Delta_c>0$ ($\Delta_c<0$), the lower (upper) band contains
more components of $|e\rangle$ states. Since the probe field only couples the ground state to the $|e\rangle$ component, the peak in the upper band gets higher compared with the other peak when we change $\Delta_c$ from positive to negative. A similar asymmetry was also predicted in four-wave-mixing in two-level atoms \cite{Friedmann1998}.
The apparent third peaks in the middle of the spectra are due to the two dips at the one- and two-photon resonances. The dip at the two-photon resonance is due to the effect of EIT. At the one-photon resonance, the power of the coupling fields is not much larger and even smaller than that of the probe field, which induces absorption of the probe field without superradiant emission \cite{supp}.

\begin{figure}
\centerline{\includegraphics[width=6cm]{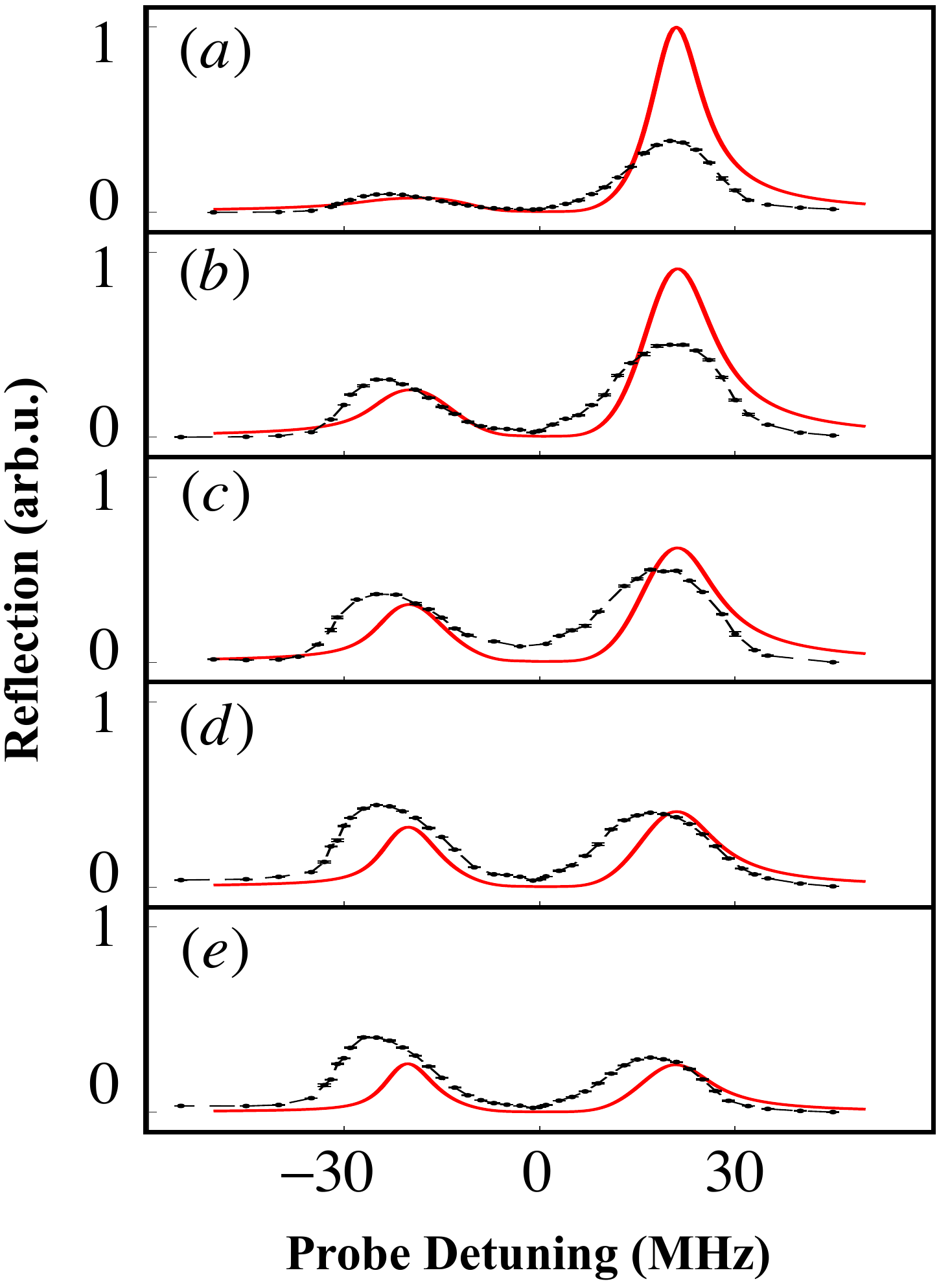}} \vspace{0.08in}
\caption{\textbf{Superradiant emission spectra for different incident angles
of the incident probe beam.} The deviation of the angle of the probe beam from $\phi=132^{o}$
are (a) $-0.6^{o}$, (b) $0^{o}$,
(c) $0.3^{o}$, (d) $0.6^{o}$ and (e)
$0.9^{o}$. The powers of each
coupling laser is 200 $\mu$W. The power of the probe light is 25 $\mu$W. $\Delta_{c}=0$. The data points are simply connective with black lines and the red curve is the theoretical fitting.
 \label{Fig4} }
\end{figure}

In the experiments, the right peak is generally higher than the left peak.
Part of the reason for the asymmetry between the two peaks is attributed to the dispersion
of the light field in BEC. The right peak can be tuned
lower than the left peak by changing the angle of the incident probe field.
This is similar to the phase matching in four-wave-mixing
\cite{Zhang2011,Kang2004}. The spectra due to different
angles of the incident probe beam is plotted in Fig. \ref{Fig4}.
Since the BEC is highly dispersive to the probe light
near the EIT point, we must take into account the phase mismatch due to this dispersion,
$\Delta k=2[n(\nu_{p})\nu_{p}\cos(\phi/2)-\nu_{c} \sin(\theta/2)]/c$,
where $\nu_{p}$ and $\nu_{c}$ are the frequencies of
the probe and coupling fields and $n$ is the refraction index of the BEC. For fixed $\phi$ and $\theta$,
the phase mismatch $\Delta k$ is basically linear to the probe detuning across
the single photon resonance of the probe field. The peak that has smaller phase
mismatch is higher than the other peak.

The asymmetry observed in the experiment cannot be fully explained by the dispersion in our current model \cite{supp}.
The two peaks are generally shifted to the low frequency side.
This can be attributed to two factors that are not taken into account in the current theoretical model. One is
the nonzero momentum states in the BEC due to the trapping
and interaction between atoms. The other is the collective decay rate and Lamb shift of the superradiant states \cite{Scully2009}.
A detailed calculation of these two quantities in the current experiment is difficult due to the near resonance condition \cite{Zhang1993,Javanainen1995}.

The superradiant emission with a standing-wave coupled three-level EIT configuration is different from
the ordinary Bragg scattering \cite{Schilke2011,Horsley2013} or the degenerate four-wave mixing \cite{Gatto2006} from an ensemble of density modulated cold atoms, where only two atomic levels are relevant. In those cases, the standing wave forms an optical lattice that
modulates the density of the atomic ensemble, such that the refractive index of the probe beam
is periodically modulated and a photonic crystal is formed. The reflection spectra demonstrate the
band structure of a photonic crystal. Although some density modulation might still exist in our experiments, the reflection spectra specifically demonstrate the band structure of a momentum-space tight-binding lattice. Each state in this BEC superradiance lattice is analogous to the timed Dicke state \cite{Scully2006} or the collective spin-wave excitation in the DLCZ protocol \cite{Duan2001}. In the Supplementary Material \cite{supp}, we have compared the results of the superradiance lattice in BEC and density modulation induced photonic crystal in BEC, as well as the results of BEC and noncondensed cold atoms.

In conclusion, we have shown the superradiance spectra of a
one-dimensional superradiance lattice with ultracold $^{87}$Rb
atoms. The relation between the spectra and the coupling field
detuning and intensity can be explained by the band structure of the
superradiance lattice. Many-body effect due to interactions between
atoms might be responsible for the asymmetries in the spectra. The
lattice can be generalized to two dimensions where the Haldane model
can be simulated \cite{Wang2015-1}.

$^*$pengjun$\_$wang@sxu.edu.cn.

$^\dagger$hcai@tamu.edu.

$^{\ddagger}$jzhang74@sxu.edu.cn,
jzhang74@yahoo.com.

\begin{acknowledgments}
This research is supported by the National Key R$\&$D Program of China (Grants No. 2016YFA0301602, No. 2016YFA0301603),
NSFC (Grants No. 11474188, 11704234), 
 the Fund for Shanxi ``1331 Project'' Key Subjects Construction,
 the program of Youth Sanjin Scholar,
 Program of State Key Laboratory of Quantum Optics and Quantum Optics Devices(No. KF 201702),
 and the Key Research Program of the Chinese Academy of Sciences (Grant No. XDPB08-3).
 H. Cai is supported by Herman F. Heep and Minnie Belle Heep Texas A$\&$M University Endowed Fund.

\end{acknowledgments}


\begin{thebibliography}{10}
\bibitem{Greiner2002} M. Greiner, O. Mandel, T. Esslinger, T. W. Hansch, and I.
Bloch, Nature \textbf{415}, 39
(2002).

\bibitem{Bloch2008} I. Bloch, J. Dalibard, W. Zwerger, Rev. Mod. Phys. \textbf{80}, 885 (2008).

\bibitem{Davis1995} K. B. Davis, M.-O. Mewes, M. R. Andrews, N. J. van Druten, D. S. Durfee, D. M. Kurn, and W. Ketterle, Phys. Rev. Lett. 75, 3969 (1995).

\bibitem{Birkl1995} G. Birkl, M. Gatzke, I. H. Deutsch, S. L. Rolston, and W. D. Phillips, Phys. Rev . Lett. 75, 2823 (1995).

\bibitem{Weidemuller1995} M. Weidemuller et al., Phys. Rev. Lett. 75, 4583 (1995).

\bibitem{Slama2005} S. Slama, C. von Cube, B. Deh, A. Ludewig, C. Zimmermann, and Ph. W. Courteille, Phys. Rev. Lett. 94, 193901 (2005).

\bibitem{Schilke2011} A. Schilke, C. Zimmermann, P. W. Courteille, and W. Guerin, Phys. Rev. Lett. 106(22), 223903 (2011).

\bibitem{Hart2015} R. A. Hart, P. M. Duarte, T. L. Yang, X. Liu, T. Paiva, E. Khatami, R. T. Scalettar, N. Trivedi, D. A. Huse, R. G. Hulet, Nature 519, 211 (2015).

\bibitem{Bajcsy2003} M. Bajcsy, A. S. Zibrov, and M. D. Lukin, Nature 426(6967), 638 (2003).

\bibitem{Bae2008} I. H. Bae, H. S. Moon, M. K. Kim, L. Lee, and J. B. Kim, Appl. Opt. 47, 4849 (2008).

\bibitem{Zhang2011} Y. P. Zhang, Z. G. Wang, Z. Q. Nie, C. B. Li, H. X. Chen, K. Q.
Lu, and M. Xiao, Phys. Rev. Lett. 106, 093904 (2011).

\bibitem{Wang2013} D. W. Wang, H. T. Zhou, M. J. Guo, J. X. Zhang, J. Evers, and S. Y.
Zhu, Phys. Rev. Lett. 110, 093901 (2013).

\bibitem{Meystre} P. Meystre, \text{Atom Optics}, Springer-Verlag 2001.

\bibitem{Dicke1954} R. Dicke, Phys. Rev. 93, 99 (1954).

\bibitem{Feld1973} N. Skribanowitz, I. P. Herman, J. C. MacGillivray, and M. S. Feld, Phys. Rev. Lett. 30, 309 (1973).

\bibitem{DeVoe1995}R. G. DeVoe and R. G. Brewer, Phys. Rev. Lett. 76, 2049 (1995).

\bibitem{Kimble2015} A. Goban, C.-L. Hung, J. D. Hood, S.-P. Yu, J. A. Muniz, O. Painter, and H. J. Kimble, Phys. Rev. Lett. 115, 063601 (2015).



\bibitem{Inouye1999} S. Inouye, A. P. Chikkatur, D. M. Stamper-Kurn, J. Stenger, and W. Ketterle. Science 285, 571 (1999).

\bibitem{Baumann2010} K. Baumann, C. Guerlin, F. Brennecke and T. Esslinger, Nature 464, 1301 (2010).

\bibitem{Dell2003} D. H. J. O'Dell, S. Giovanazzi, and G. Kurizki, Phys. Rev. Lett. 90, 110402 (2003).

\bibitem{Zhang1993} W. Zhang, Phys. Lett. A 176, 225 (1993).

\bibitem{Steinhauer2002} J. Steinhauer, R. Ozeri, N. Katz, and N. Davidson Phys. Rev. Lett. 88, 120407 (2002).

\bibitem{Araujo2016} M. O. Araujo, I. Kresic, R. Kaiser, and W. Guerin, Phys. Rev. Lett. 117, 073002 (2016).

\bibitem{Roof2016} S. J. Roof, K. J. Kemp, M. D. Havey, and I. M. Sokolov Phys. Rev. Lett. 117, 073003 (2016).


\bibitem{Scully2006} M. O. Scully, E. S. Fry, C. H. Raymond Ooi, and K. Wodkiewicz, Phys. Rev. Lett. 96, 010501 (2006).

\bibitem{Bienaime2012} T. Bienaime, N. Piovella and R. Kaiser, Phys. Rev. Lett. 108, 123602 (2012).

\bibitem{Scully2015} M. O. Scully, Phys. Rev. Lett. 115, 243602 (2015)

\bibitem{Bromley2016} S. L. Bromley, B. Zhu, M. Bishof, X. Zhang, T. Bothwell, J. Schachenmayer, T. L. Nicholson, R. Kaiser, S. F. Yelin, M. D. Lukin, A. M. Rey, and J. Ye, Nat. Commun. 7, 11039 (2016).

\bibitem{Guerin2016} W. Guerin, M. O. Araujo, and R. Kaiser, Phys. Rev. Lett. 116 083601 (2016)


\bibitem{Scully2009} M. O. Scully, Phys. Rev. Lett. 102, 143601 (2009).

\bibitem{Rohlsberger2010} R. R\"ohlsberger, K. Schlage, B. Sahoo, S. Couet and R. R\"uffer, Science 328, 1248 (2010).

\bibitem{Keaveney2012} J. Keaveney, A. Sargsyan, U. Krohn, I. G. Hughes, D. Sarkisyan and C. S. Adams, Phys. Rev. Lett. 108, 173601 (2012).

\bibitem{Meir2014} Z. Meir, O. Schwartz, E. Shahmoon, D. Oron and R. Ozeri, Phys. Rev. Lett. 113, 193002 (2014).

\bibitem{Wang2015} D. W. Wang, R. B. Liu, S.-Y. Zhu and  M. O. Scully, Phys. Rev. Lett. 114 043602 (2015).

\bibitem{supp} Supplementary material for the experimental set-up and theoretical calculation and fitting for the experimental data, which includes
Refs. \cite{Price, PJWang, Cheuk, Dahan, Polkovnikov, Hooley, Lin, Andre, Adams}.

\bibitem{Price} H. M. Price, T. Ozawa, and I. Carusotto, Phys. Rev. Lett. 113 190403 (2014).

\bibitem{PJWang} P. Wang, Z.-Q. Yu, Z. Fu, J. Miao, L. Huang, S. Chai, H. Zhai, and J. Zhang, Phys. Rev. Lett.109 095301 (2012).

\bibitem{Cheuk} L. W. Cheuk, A. T. Sommer, Z. Hadzibabic, T. Yefsah, W. S. Bakr, and M. W. Zwierlein, Phys. Rev. Lett. 109 095302 (2012).

\bibitem{Dahan} M. B. Dahan, E. Peik, J. Reichel, Y. Castin, and C. Salomon, Phys. Rev. Lett. 76, 4508 (1996).

\bibitem{Polkovnikov} A. Polkovnikov, S. Sachdev, and S. M. Girvin, Phys. Rev. A 66 053607 (2002).

\bibitem{Hooley} C. Hooley and J. Quintanilla, Phys. Rev. Lett. 93 080404 (2004).

\bibitem{Lin} Y.-J. Lin, K. Jim{\'{e}}nez-Garc{\'{\i}}a, and I. B. Spielman, Nature 471, 83 (2011).

\bibitem{Andre} A. Andr\'e and M. D. Lukin, Phys. Rev. Lett. 89, 143602 (2002).

\bibitem{Adams} B. W. Adams, Journal of Modern Optics, 56, 1974 (2009).

\bibitem{Friedmann1998}H. Friedmann and A. D. Wilson-Gordon, Phys. Rev. A 57, 4854 (1998).

\bibitem{Kang2004} H. Kang, G. Hernandez, and Y. Zhu, Phys. Rev. A 70, 061804 (2004).

\bibitem{Javanainen1995} J. Javanainen, Phys. Rev. Lett. 75, 1927 (1995).

\bibitem{Horsley2013} S. A. R. Horsley, J. H. Wu, M. Artoni, and G. C. La Rocca, Phys. Rev. Lett. 110, 223602 (2013).

\bibitem{Gatto2006} G. L. Gattobigio, F. Michaud, J. Javaloyes, J. W. R. Yabosa, and R. Kaiser,
Phys. Rev. A 74, 043407 (2006).

\bibitem{Duan2001} L. M. Duan, M. D. Lukin, J. I. Cirac and P. Zoller, Nature 414, 413 (2001).

\bibitem{Wang2015-1} D. W. Wang, H. Cai, L. Yuan, R. B. Liu, S.-Y. Zhu and  M. O. Scully, Optica 2, 712 (2015).






\end{thebibliography}
\end{document}